\documentclass[twocolumn,pre]{revtex4}  
\usepackage{graphicx}
\usepackage{dcolumn}
\usepackage{bm}
\usepackage{amsxtra}
\usepackage{latexsym}
\usepackage{comment}
\usepackage{subfigure}

\newcommand{\ket}[1]{\mbox{\ensuremath{ | #1 \rangle }}}                           

\newcommand{\inprod}[2]{\mbox{\ensuremath{ \langle #1 | #2 \rangle}}}

\begin{document}

\preprint{The contact process in disordered and periodic binary
  two-dimensional lattices}  

\title{The contact process in disordered and periodic binary two-dimensional lattices}

\author{S.~V.~ Fallert}
 \email{sf287@cam.ac.uk}  
\affiliation{Department of Chemistry, University of Cambridge,  
             Cambridge, UK }

\author{Y.~M.~ Kim}  
\affiliation{St. Catharine's College, 
University of Cambridge, Cambridge, UK}
  
\author{C.~J.~ Neugebauer}  
\affiliation{Department of Chemistry, University of Cambridge, Cambridge, UK}

\author{S.~N.~Taraskin}  
\affiliation{St. Catharine's College and Department of Chemistry, University of Cambridge,  
             Cambridge, UK}  

\date{\today}
  
\begin{abstract} 
The critical behavior of the contact process in disordered and
periodic binary $2d$-lattices is investigated numerically by means of
Monte Carlo simulations as well as via an 
analytical approximation and standard mean field theory. 
Phase-separation lines calculated numerically are found to agree well 
with analytical predictions around the homogeneous point.
For the disordered case, values of static scaling exponents obtained
via quasi-stationary simulations are found to change with disorder
strength.
In particular, the finite-size scaling exponent of the density of
infected sites approaches a value consistent with the existence of an
infinite-randomness fixed point as conjectured before for the $2d$
disordered CP.
At the same time, both dynamical and static scaling exponents are
found to coincide with the values established for the homogeneous 
case thus confirming that the contact process in a heterogeneous 
environment belongs to the directed percolation universality
class.
\end{abstract}  
  
\pacs{05.70.Ln,64.60.Ht,02.50.Ey,87.18.Bb}

  
\maketitle



\section{\label{sec:intro} Introduction}

The contact process (CP) \cite{Harris_74}, is a prototype model for 
the spatial spread of epidemics in biological systems.
It describes
 epidemics in populations where each member can be in one of two states:
infected (I) or susceptible (S) (so-called SIS models). 
The CP exhibits a non-equilibrium phase transition between an active and
a non-active regime of the disease, behaving at its critical point according 
to the directed percolation (DP) universality class.
This has been established by a 
range of analytical and numerical techniques 
\cite{Liggett_85:book,Marro_99:book,Hinrichsen_00:review,Odor_04:review} 
such as renormalization group analysis 
\cite{hooyberghs_03,Odor_04:review}, 
series expansions \cite{jensen_93}, 
Monte Carlo (MC) simulations \cite{Grassberger_79,Grassberger_89}
 and spectral analysis of the 
Liouville operator \cite{Mendonca_99,deoliveira2006}. 
These analyses have been undertaken for simple topologies, 
mostly for homogeneous hyper-cubic lattices. 

Recently, interest has turned towards the behavior of this process in
disordered environments and revealed very peculiar features
such as changing exponents and significantly different
dynamics like Griffiths phases and activated scaling
\cite{dickman_96,dickman_98,Vojta_05}.
In general, heterogeneous environments are typical in realistic 
systems,  especially in the context of control
of epidemics \cite{Finckha_99,Zhu_00,Otten_05,forster_07}.   
Therefore, it is instructive to investigate the critical 
behavior of the CP under these conditions and in particular to   
establish the phase diagrams for such systems. 
In the past, the disordered CP (DCP) has been investigated in both one
and two dimensions in a range of settings and revealed a continuous
change in static critical exponents starting from the clean DP values
\cite{dickman_98,hooyberghs_04,Neugebauer_06}.
In the $1d$ case, a strong-disorder renormalization group study allowed
deep insight into the disordered process and revealed
a dominating infinite-randomness fixed point (IRFP) for sufficiently
strong disorder \cite{hooyberghs_03}.
In the weak- to intermediate-disorder regime however, MC
simulations and density-matrix renormalization group (DMRG) techniques
\cite{hooyberghs_04} as well as series expansions \cite{Neugebauer_06}
found continuously varying disorder-dependent critical exponents which
were found to approach those characteristic of an IRFP with increasing
strength of disorder.
For the $2d$ CP with site dilution, MC simulations showed a similar
behavior, a continuous change in exponents with increasing disorder
\cite{dickman_96,dickman_98} and, in
retrospect \cite{hooyberghs_04}, giving evidence for the existence of an
IRFP also in this case.

In this paper, we consider the phase diagram of the CP in a binary $2d$
lattice of sites with different recovery rates
$\epsilon_A$ and $\epsilon_B$ drawn from
a bimodal probability distribution.
Extensive Monte Carlo (MC) simulations following \cite{dickman_96} are
employed  in order to
locate the line of critical points in the space of recovery rates
$(\epsilon_A,\epsilon_B$).
As such simulations of disordered systems are numerically intensive
due to very long relaxation times, analytical approximations
are vital to constrain the region of phase space which
contains the line of critical points.
Here, we analyze the results obtained from the mean field
approximation and further propose an approximate expression for
critical points guided by the structure of the Liouville operator
which governs the time evolution of the CP.

Also, the quasi-stationary (QS) simulation method \cite{oliveira_2005} 
is employed to
investigate the static scaling behavior of the CP in this disordered
system and to calculate the corresponding critical exponents.
In particular, we study whether the process in this setting
exhibits disorder-dependent changing exponents which cross over to
values characteristic of an IRFP for sufficiently strong disorder as
observed previously \cite{hooyberghs_04,dickman_98}.

Following on, we investigate the behavior of the CP in
a range of heterogeneous periodic lattices with different unit cells 
via MC simulations and test the 
validity of our analytical expression as well as standard mean 
field theory.
The two analytical approaches, mean field and our
alternative approximation, enable us to largely constrain the location
of the critical points in both the disordered and the heterogeneous
periodic lattices.
Critical exponents are found to change continuously in the former case
with increasing disorder and appear to approach the predicted values
characteristic of an IRFP while they remain constant at their DP
values in the latter.

The CP, its critical behavior and some of the theoretical
foundations employed for its description are introduced in Sec.~\ref{sec:bg}. 
Our analysis of the disordered system is presented in
Sec.~\ref{sec:disorder}.
In Sec.~\ref{sec:hetero} we investigate the CP in a range of
heterogeneous periodic lattices in a similar fashion.
Lastly, our findings are discussed in Sec.~\ref{sec:discussion} and we
summarize in Sec.~\ref{sec:conclusion}.


\section{\label{sec:bg} Background}

In this section, we define the CP and give an overview of its
critical behavior and the master-equation description by means of the
Liouville operator. 
The CP is a non-equilibrium stochastic process in which an infection 
spreads via nearest-neighbor contact from site $i$ to $j$ at a transmission 
rate $w_{i \to j}$. 
Recovery of site $i$ is spontaneous and happens at a recovery 
rate $\epsilon_i$. 
In the thermodynamic limit, 
the ratio of these two rates is the control parameter of  
a second-order phase transition  between a non-active phase where no
infected sites remain as $t \to \infty$ and an active phase where the 
density of infected sites (order parameter) is non-zero  as $t \to \infty$ 
\cite{Liggett_85:book,Hinrichsen_00:review,Marro_99:book}.

For the CP in a system of size $N$ with sites $i=1 \ldots N$ we denote
the two possible
states of site $i$ as $s_i=1$ (infected) or $s_i=0$ (susceptible). 
A microstate of the system, i.e. a snapshot of the infection states of
all sites, can be defined as a vector  $\mathbf{S}= (s_1,
\ldots, s_N)^T$ and the probability of finding the system in a
specific microstate at time $t$ is denoted by $P(\mathbf{S},t)$. 
Assuming the transition rates between microstates 
$\mathbf{S}$ and $\mathbf{S}'$ to be $r_{\mathbf{S} \rightarrow \mathbf{S}'}$,
the time evolution of this probability follows the master equation
which expresses the conservation of probability flow,
\begin{eqnarray}
\partial_t P(\mathbf{S},t) = \sum_{\mathbf{S}'} \left( r_{\mathbf{S}'
    \rightarrow \mathbf{S}} P(\mathbf{S}',t) - r_{\mathbf{S} \rightarrow \mathbf{S}'} P(\mathbf{S},t) \right)~,
\label{eq:master_equation}
\end{eqnarray}
where the transition rates $r_{\mathbf{S} \rightarrow \mathbf{S}'}$ follow
from the rules of the CP.
The master equation can be recast in compact form by introduction of
the Liouville operator $\hat{\mathcal L}$ which acts on the
probability state vector
$\ket{P(t)}$,
\begin{equation}
\partial_t \ket{P(t)} = \hat{\mathcal L} \ket{P(t)}~,
\label{e1}
\end{equation}
the components of which are the probabilities 
of finding a system of $N$ sites in different states $\ket{\sigma}$ at time $t$,  
$\ket{P(t)} = \sum_{\sigma} \inprod{\sigma} {P(t)} \ket{\sigma}$.
Here, $\left\{ \ket{\sigma} \right\}$ is the orthonormal basis diagonal in
the occupation number representation \cite{Marro_99:book,Mendonca_99}.
The precise form of $\hat{L}$ is most readily expressed in terms of
hard-core bosonic creation and annihilation operators acting on site
$i$, $a_i^{\dagger}$ and $a_i$, respectively,
\begin{eqnarray}
\hat{L} 
         =   \sum_i \left( \epsilon_i (1-a^{\dagger}_i) a_i +
           (1-a_i)a^{\dagger}_i \sum_{j \in NN(i)} w_{j \to i}
           ~ a_{j}^{\dagger} a_{j} \right)~,
\end{eqnarray}
%
%
where the first part destroys particles while the second part creates
offspring \cite{Marro_99:book}. 
The Liouville operator is non-Hermitian with 
matrix elements $\hat{\mathcal L}_{\sigma' \sigma} \equiv \langle
\sigma'|\hat{\mathcal L}| \sigma \rangle$, which  coincide  with the
transition 
rates from state $\sigma$ to state $\sigma'\ne \sigma$ and  
$\hat{\mathcal L}_{\sigma \sigma} =-\sum_{\sigma'\ne \sigma}
\hat{\mathcal L}_{\sigma' \sigma}$. 

In principle, Eq.~(\ref{e1}) can be solved by performing direct
diagonalization of the $2^N\times 2^N$ real sparse (for lattice
topologies) non-symmetric Liouville matrix. 
Its formal solution can then be expressed as
\begin{eqnarray}
\ket{P(t)} & = & \sum_i \textrm{e}^{\lambda_i t} \inprod{e_i}{P(0)} ~ \ket{e_i}~,
\label{eq:time_evolution}
\end{eqnarray}
where $\lambda_i$ are the eigenvalues of the Liouville matrix with a complete set
of eigenstates $\ket{e_i}$. 
The trivial solution $\ket{e_0}$ of the eigenproblem for the Liouville
operator with $\lambda_0=0$ corresponds to the absorbing state of the system. 
All other eigenvectors $\ket{e_i}$ in finite systems 
have eigenvalues with negative real parts and thus decay exponentially 
with time. 
In the thermodynamic limit ($N\to \infty$), there is one eigenstate
$\ket{e_1}$ with corresponding eigenvalue, $\lambda_1$, which is zero in the
active and non-zero in the non-active phase. 
In a finite system, 
the value of $\lambda_1$ in the active (non-active) regime approaches 
a zero (non-zero) value with increasing $N$, thereby signaling
the phase transition. 
The exact
location of the transition can be extrapolated using finite-size data for
moderate system sizes (e.g.\ $N\leq 16$) from direct diagonalization or 
density-matrix renormalization group calculations 
\cite{hooyberghs_03,Mendonca_99}.


\section{\label{sec:disorder} Disordered System}

In what follows, we investigate the behavior of the CP on a lattice of
two types of site, $A$ and $B$, characterized by different recovery
rates $\epsilon_A$ and $\epsilon_B$, respectively.
The recovery rate at site $i$, $\epsilon_i$, is drawn from the bimodal distribution
\begin{equation}
p(\epsilon_i) = x~\delta(\epsilon_i-\epsilon_A) +(1-x)~\delta(\epsilon_i-\epsilon_B)~,
\label{eq:bimodal}
\end{equation}
where $x$ controls the relative concentration of $A$ and $B$ sites.
The transmission rate, for simplicity,  
is the same for all possible links between nodes, 
$w_{i \to j} = w_{j \to i} =w$.    
As a further simplification, the timescale is set up by choosing
$w=1/Z$ with $Z$ being the number of nearest-neighbor links per 
node ($Z=4$ for the topologies considered here).

\subsection{\label{sec:disorder_mc}MC Simulation}

\begin{figure}[ht] %
\begin{center}
\scalebox{0.32}{\includegraphics[angle=270]{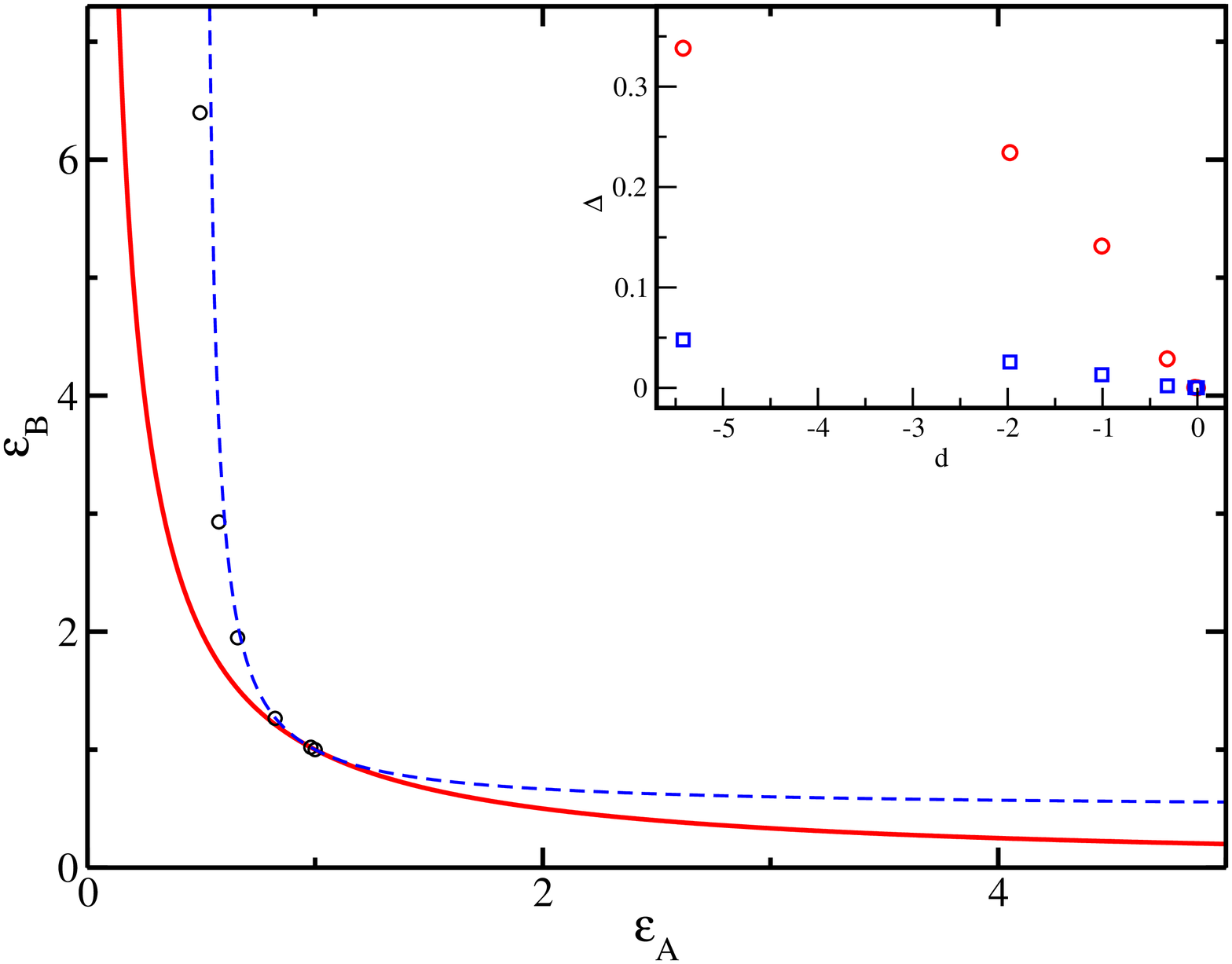}}
\end{center}
\caption{(Color online) The phase diagram for the CP on a disordered
  lattice with sites of recovery rates $\varepsilon_A$ or
  $\varepsilon_B$ drawn from the distribution Eq.~(\ref{eq:bimodal}) for the
case $x=0.5$ obtained from MC simulation (dots), mean field (upper dashed line) and the analytical expression
Eq.~(\ref{eq:locus}) (solid line). Inset shows the deviation $\Delta(d)$
as defined in the text for both mean field (blue $\Box$) and
Eq.~(\ref{eq:locus}) (red $\circ$).}
\label{fig:disorder}
\end{figure}

MC simulations are used to locate the critical point by starting from
a single infection seed and averaging over $10^6$ realizations of the 
process each with a fresh
realization of the disorder up to a maximum of $10^6$ time steps.
We consider the case of $x=0.5$ and a range of values for
$\epsilon_A$, aiming to find the corresponding critical $\epsilon_B$
for each.
As previously observed \cite{dickman_96}, very slow dynamics are
encountered, an
effect that increases with disorder strength, i.e. the distance 
between $\epsilon_A$ and the homogeneous critical rate 
$\epsilon_c = 0.60653(3)$ \cite{Marro_99:book}.
Following \cite{dickman_96}, the criterion of asymptotic monotonic
growth or decay in order to assess whether the process is super- or
sub-critical for a particular choice of rates is employed.
For clarity of presentation, rescaled recovery rates 
$\varepsilon_i \equiv \epsilon_i / \epsilon_c$ are introduced in order 
for the homogeneous critical point to be conveniently located at 
$\varepsilon_c=\varepsilon_A = \varepsilon_B = 1$.
The resulting phase diagram, symmetric in $\varepsilon_A$ and
$\varepsilon_B$, is shown in Fig.~\ref{fig:disorder}.

\subsection{\label{sec:disorder_mf}Mean Field Approximation}

As outlined in the introduction, we are interested in analytically 
approximating the region where the phase separation line is
located.
To this end, we first present an approach based on mean field theory
\cite{Marro_99:book}.
In this approximation, fluctuations and correlations are ignored
rendering the master equation analytically tractable.
For the case of the disordered system considered above, the governing
equations for the mean concentrations of infected sites of type $A$
and $B$, $n_A$ and $n_B$ respectively, are given by
\begin{eqnarray}
\frac{\partial n_A}{\partial t} &=&
     -\epsilon_A n_A + \frac{\epsilon_*}{2} (1-x) n_B(1 - n_A) 
                     + \frac{\epsilon_*}{2} x n_A(1 - n_A)
     \nonumber \\
\frac{\partial n_B}{\partial t} &=&
     -\epsilon_B n_B + \frac{\epsilon_*}{2} x n_A(1 - n_B) 
                     + \frac{\epsilon_*}{2} (1-x) n_B(1 - n_B) ~,
\label{eq:disorder_mf} 
\end{eqnarray}
where $\epsilon_*=wZ=1$ is the mean field critical value for the 
recovery rate in the homogeneous $2d$ square lattice. 
As usual for the mean field approximation in low-dimensional systems, 
$\epsilon_*$ significantly overestimates the true critical value 
for the homogeneous case, $\epsilon_A = \epsilon_B$. 
The locus of critical points in the parameter space ($\epsilon_A, \epsilon_B$) 
separating non-active and active phases can be easily found from the 
solution of Eqs.~(\ref{eq:disorder_mf}) in the steady-state regime giving
\begin{eqnarray}
\left\langle \frac{1}{\varepsilon_i} \right\rangle = 1
\label{eq:mf_disorder}
\end{eqnarray}
with $\varepsilon_{i}=\epsilon_{i}/\epsilon_*$ where $\langle \dots
\rangle$ denotes an average over disorder realizations.

The resulting phase separation line is shown in
Fig.~\ref{fig:disorder} along with the MC data presented above.
Note that due to rescaling, mean field and numerical results coincide
by construction at the homogeneous critical point.
In order to allow a quantitative comparison between numerical results
and approximation, a measure of difference between prediction and the
true value obtained by MC simulation is needed.
As such a measure, we consider the shortest distance
$\Delta(d)$ between the prediction curve and an MC data point
a (shortest path) distance $d$ away from the homogeneous point.
This error quantity is suitable for quantitative analysis as it
is a measure for the width of the region of uncertainty between the
analytical prediction and the true critical line and will be symmetric
about the homogeneous point for symmetric phase diagrams.
The inset of Fig.~\ref{fig:disorder} shows $\Delta(d)$ for the mean field
approximation (blue squares).

As can be seen from the figure, the mean field approximation provides an upper
bound to a region that contains the phase separation line.
While the deviation $\Delta(d)$ is small in the vicinity of the
homogeneous critical point ($\Delta < 0.01$), it grows considerably as
the degree of heterogeneity increases ($\Delta \approx 0.1$).

\subsection{\label{sec:disorder_analytic}Alternative Analytical Approximation}

Given that the mean field approximation appears to provide an upper
bound to the region which contains the phase separation line, we are
interested in obtaining an alternative analytical approximation that
may provide a lower bound.
In the following we will first present an approximate expression for
the location of critical points in a heterogeneous system and then
compare its predictions to the MC data of section
\ref{sec:disorder_mc}.
Following on, we give a motivation for this approximation along with
numerical support. 

\subsubsection*{Statement and Comparison to Data}

Consider a finite system of $N$ sites with arbitrary recovery
rates $\epsilon_i$ ($i=1,\dots,N$).
We will argue below that for such a system in the vicinity of the
homogeneous critical point (all $\epsilon_i=\epsilon_c$) the expression
\begin{equation}
\prod_{j}^{N}\epsilon_{j} = \epsilon_c^N
~,    
\label{eq:locus}
\end{equation}
approximately predicts the location of critical points.
For the disordered system presented earlier, this expression
simplifies to $\epsilon_c^2 = \epsilon_A~\epsilon_B$.
Fig.~\ref{fig:disorder} shows both the MC data presented earlier and
the approximate line of critical points thus obtained.
As can be seen from the figure, the alternative
analytical approximation is found to provide a reliable lower bound to the
region which contains the line of critical points.
Hence, in combination with the mean field approximation discussed above, one
can constrain this region.
Considering the error $\Delta(d)$, it is found to show similar
behavior to the one previously observed for the mean field data albeit an 
order of magnitude larger.

\subsubsection*{Motivation and Numerical Support}

In the following we motivate Eq.(\ref{eq:locus}) by considering the
structure of the Liouville operator as defined in Sec.~\ref{sec:bg}.
For the finite system introduced above, the eigenvalues of the Liouville
operator, $\lambda$, are given by the characteristic equation,
\begin{equation} 
\lambda^{-1}|\lambda\hat{I} -\hat{\mathcal L}|
=
Q_{N_{\text{max}}}(\{ \epsilon_i \},\lambda)
= 
\sum_{n=0}^{N_{\text{max}}} A_n(\{ \epsilon_i \}) \lambda^n = 
0~,   
\label{eq:Eq1}
\end{equation}
where $Q_{N_{\text{max}}}(\lambda)$ 
is a polynomial in $\lambda$ of order $N_{\text{max}}$ 
($N_{\text{max}}=2^{N}-1$) and   
division by $\lambda$ eliminates the trivial zero root for 
the absorbing state.
It is our aim to solve this equation approximately in the vicinity of
the homogeneous critical point where  $\epsilon_i
= \epsilon_c$.

To this end, we first consider the coefficients $A_n$ and look for
features in their structure which may help in rendering the equation
tractable. 
Generally, the $A_n$ can be expressed as
\begin{equation} 
 A_n(\{ \epsilon_i \}) = \sum_{m_1,\ldots ,m_N=0} \alpha^{(n)}_{m_1, \ldots,
  m_N} ~ \prod_{j}^{N}\epsilon_{j}^{m_j}~,   
\label{eq:A_n}
\end{equation}
where  the upper limits in the sum depend on $n$ but their precise values are
not significant for the analysis below. 
We now assume that in the construction of the $A_n(\{ \epsilon_i \})$
from the determinant of $\lambda\hat{I} -\hat{\mathcal L}$, the
dominant contribution stems from terms with 
products of the \emph{same} (or at least similar) powers of recovery
rates at different sites.
If this is true, the previous equation can be approximated as
\begin{equation}
A_n(\{ \epsilon_i \}) 
\simeq  
A_n \bigl(\prod_{j}^{N}\epsilon_{j} \bigr)
= 
\sum_{m=0}^{m_*} \alpha_{m}^{(n)} 
~ \left(\prod_{j}^{N}\epsilon_{j}\right)^{m} ~,    
\label{eq:F_approx}
\end{equation}
where $m_* \alt 2^N/N$.
While this assumption may at first appear artificial, a justification
can be found in the structure of the Liouville operator.
The determinant of the Liouville matrix contains the sum of terms which are 
the products of the recovery rates (the transmission rates are chosen to be
constant).
A typical (representative) term contains a product of many  
recovery rates, each one picked from a different column.
Assuming periodic boundary conditions, all sites in the system should
enter the Liouville matrix in the same fashion.
Therefore, in a typical term one would not expect to find an
over-representation of a specific site leading to the statement that
the (combinatorially) dominant terms correspond to products of
recovery rates raised to powers that are close in value.

This argument only holds if one can be sure that the combinatorial
weight of terms with homogeneous powers is not offset by the actual
values of the recovery rates $\epsilon_i$.
Otherwise, one could imagine the dominance of terms with very
different powers of $\epsilon_i$ caused by the raising of values $>1$ 
to a high power.
However, recall that for the clean CP at the homogeneous critical
point, the true critical
recovery rate is $<1$.
Thus, close to this point, the critical recovery rates $\epsilon_i$ will
always be close in value and $<1$ which means that the above
argument about homogeneous powers is expected to hold in this regime.
Further support will be given below in the form of numerical evidence 
using a specific system further down.

As explained in Sec.\ref{sec:bg}, at criticality the highest
non-trivial eigenvalue $\lambda_1(N)$ will be finite and tends to zero
with increasing $N$.
For the case of homogeneous recovery rates at criticality, $\epsilon_i
= \epsilon_c$ and $Q_{N_{\text{max}}}(\{ \epsilon_c \},\lambda_1)=0$. 
Finally, by combination of this property and Eq.~(\ref{eq:F_approx}), we 
indeed find $\prod_{j}^{N}\epsilon_{j} = \epsilon_c^N$ for the homogeneous
critical point which is precisely the statement presented above in
Eq.~(\ref{eq:locus}). 
 
The last step of our approximation then is to employ the same relation
away from the homogeneous point and to use it to predict the locus of
critical points.

More formally, the above condition for critical points can be derived from 
Eqs.~(\ref{eq:Eq1}) and (\ref{eq:F_approx}) via a Taylor series expansion 
of $Q_{N_{\text{max}}}(\{ \epsilon_i \},\lambda_1)$ 
around the 
homogeneous critical point in $\ln(\epsilon_i/\epsilon_c)$,
\begin{eqnarray}
&\ &Q_{N_{\text{max}}}(\{ \epsilon_i \},\lambda_1) = 
\sum_{n=0}^{N_{\text{max}}} A_n(\{ \epsilon_i \}) \lambda_1^n
\nonumber
\\
&=&
\sum_{n=0}^{N_{\text{max}}} \left( \sum_{m=1}^{m_*}
  \alpha^{(n)}_{m}e^{m\sum_i^N \ln\epsilon_i} \right) \lambda_1^{n}
\nonumber
\\
&=&
\sum_{n=0}^{N_{\text{max}}}  \sum_{m=1}^{m_*}\alpha^{(n)}_m \epsilon_c^{mN} 
\sum_{k=1}^{\infty} \frac{m^k}{k!} \ln^k
\left(\prod_{i=1}^{N}\frac{\epsilon_i}{\epsilon_c}\right)
\lambda_1^{n} = 0
~.      
\label{eq:derivation}
\end{eqnarray}
Here, we used the relation $Q_{N_{\text{max}}}(\{ \epsilon_c \},\lambda_1)=
\sum_{n=0}^{N_{\text{max}}} A_{n}(\epsilon_c^N) \lambda^{n}_1 =0 $
which leads to no constant term in the expansion and allows
factorization of the above expression, i.e. 
\begin{equation}
Q_{N_{\text{max}}}(\{ \epsilon_i \},\lambda_1) = S ~
\ln\left(\prod_{i=1}^{N}\frac{\epsilon_i}{\epsilon_c}\right) = 0 
~, 
\label{eq:new}
\end{equation}
where
\begin{equation}
S= \sum_{n=0}^{N_{\text{max}}}  \sum_{m=1}^{m_*}\alpha^{(n)}_m \epsilon_c^{mN} 
\sum_{k=1}^{\infty} \frac{m^k}{k!} \ln^{k-1}
\left(\prod_{i=1}^{N}\frac{\epsilon_i}{\epsilon_c}\right) \lambda_1^{n}
~.
\label{eq:S}
\end{equation}
Eq.~(\ref{eq:new}) is obeyed if 
\begin{equation}
\ln\left(\prod_{i=1}^{N}\frac{\epsilon_i}{\epsilon_c}\right)=0
~,
\label{eq:locus_1}
\end{equation}
because $S\ne 0$ for arbitrary choice of $\epsilon_i$, which coincides with
the condition given by Eq.~(\ref{eq:locus}). 
Alternatively, our approximate expression can be recast as an expectation
value of logarithms \cite{Neugebauer_06}, 
\begin{equation}
E \left[ \ln \frac{\epsilon}{\epsilon_c} \right] = 0
~. 
\label{eq:locus_2}
\end{equation}
Note that this procedure  amounts to simply geometrically averaging the
recovery rates and inserting them into the clean theory.
Interestingly, the logarithm of rates well known from renormalization
group analyses of
the DCP and the random transverse-field Ising model \cite{Odor_04:review} arises naturally in our scheme. 

\begin{figure}[ht] %
\begin{center}
\scalebox{0.32}{\includegraphics[angle=270]{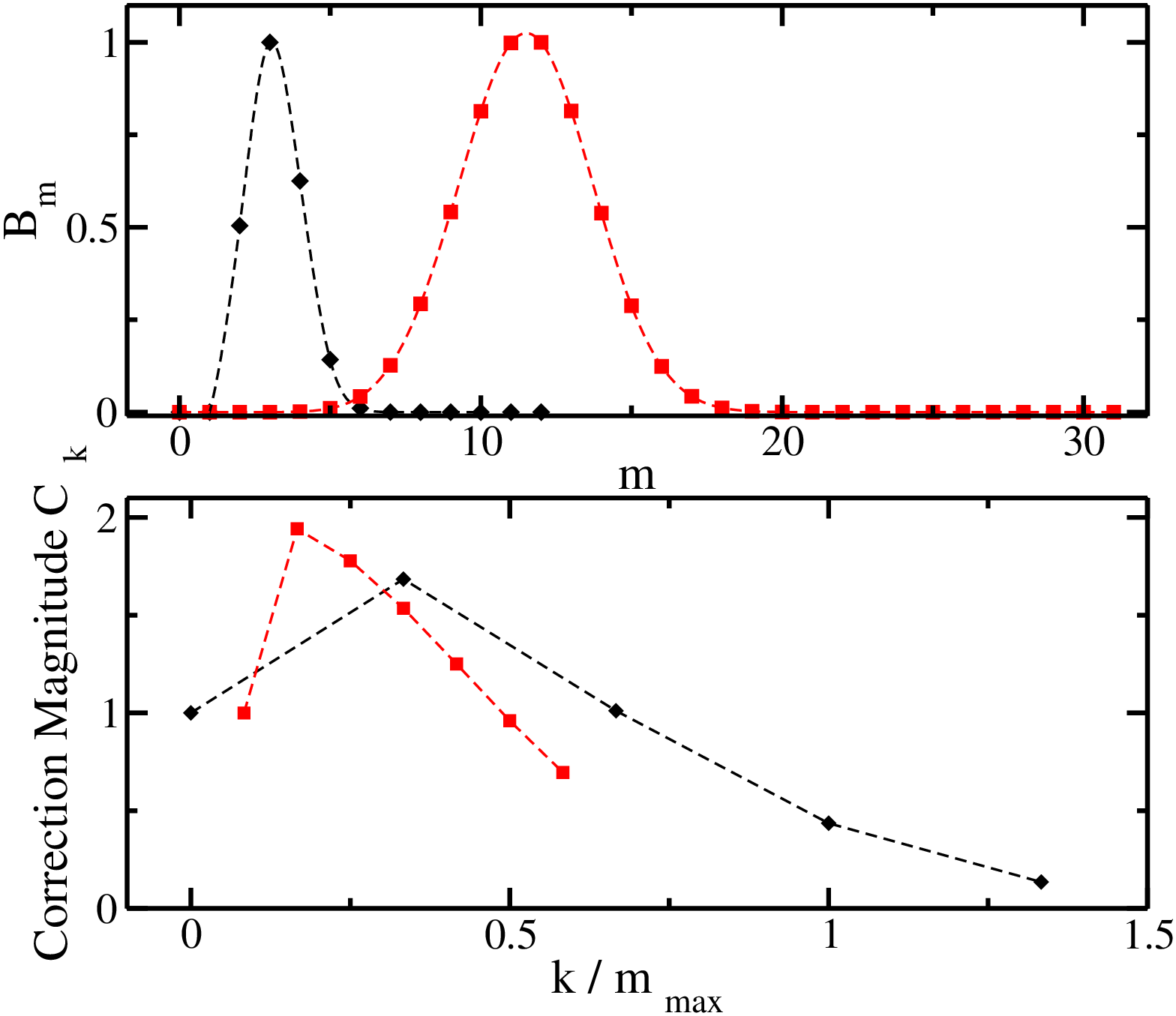}}
\end{center}
\caption{(Color online) Top panel: The terms $B_m$ as defined in
   Eq.~(\ref{eq:A_0}) for a linear $AB$ chain of size $N=4$
   (left black peak, $\Diamond$) and $N=6$ (right red peak, $\Box$)
   normalized by their maxima. Bottom panel: The correction terms $C_k$
   as defined in
   Eq.~(\ref{eq:C_k}) as a function of the relative difference between
   powers $k/m_{max}$ where $m_{max}$ is the location of the respective
   maximum in the upper panel. Symbols as before, all values have been
   normalized to the corresponding homogeneous contributions $C_0$. }
\label{fig:coefficients}
\end{figure}

In order to support the assumption about a dominant contribution to
Eq.~(\ref{eq:A_n}) from products of homogeneous powers of recovery
rates, numerical evidence for a simple system is given below.
Let us consider a $1d$ binary chain of sites $A$ and $B$ characterized 
by recovery rates $\epsilon_A$ and $\epsilon_B$, respectively, 
and spatially arranged as
$\ldots ABAB \ldots$ with periodic boundary conditions.
As a particular example, we analyze the coefficient
$A_0(\epsilon_A,\epsilon_B)$ defined by Eq.~(\ref{eq:A_n}), which reads  
(where $\alpha \equiv \alpha^{(0)}$)
\begin{eqnarray}
A_0
&=& 
\sum_{m_A,m_B} \alpha_{m_Am_B} ~
\epsilon_{A}^{m_A} \epsilon_{B}^{m_B}  
\nonumber
\\
&=&
\sum_{m=0}^{m_*} B_m(\epsilon_A,\epsilon_B)
\label{eq:A_0}
\end{eqnarray}
with 
\begin{eqnarray}
&\ & B_m(\epsilon_A,\epsilon_B) = \left( \epsilon_A \epsilon_B \right)^{m}
\times  
\nonumber
\\
&~& 
  \left(
  \alpha_{mm} + \sum_{j=1}^{m_*-m}  \alpha_{m+j,m} \epsilon_A^{j} +
  \alpha_{m,m+j} \epsilon_B^{j}   \right)
~,
\label{eq:B_n}
\end{eqnarray}
which can be symbolically evaluated for relatively small systems ($N \leq
6$). 
Initially, $\epsilon_A$ and $\epsilon_B$ will both be set equal to
$\epsilon_c$ consistent with our assumption that we investigate the
vicinity of the homogeneous critical point.
This enables us to investigate the 
relative magnitude of terms corresponding to different arrangements of powers.
The terms $B_m$ effectively correspond to the contributions which contain 
either the homogeneous power $m$ or one recovery rate to the power $m$ 
along with the other recovery rate to a power greater than $m$.
The magnitudes of the $B_m$ as  functions of $m$ for the binary system of 
size $N=4$ and $N=6$ are shown in Fig.~\ref{fig:coefficients} (top panel).
For both cases we observe sharp peaks centered at 
$m_{\text{max}}=3$ ($N=4$) and $m_{\text{max}}=12$ ($N=6$) indicating 
a dominant contribution from a narrow range of powers.

The contribution to $A_0$ from purely homogeneous powers can be written as
$C_0=\sum_{m=0}^{m_*} \alpha_{m,m} \left (\epsilon_A \epsilon_B \right )^m$
while corrections to this can be expressed as
\begin{equation}
C_k=\sum_{m=0}^{m_*-k} \left( \alpha_{m+k,m} ~ \epsilon_A^{m+k} \epsilon_B^{m} +
  \alpha_{m,m+k} ~ \epsilon_A^{m} \epsilon_B^{m+k}  \right)
\label{eq:C_k}
\end{equation}
for $k>1$. 
The values of  $C_k$ represent contributions from powers differing by $k$
from each other thus allowing a systematic investigation of the
validity of our assumption.
We are interested in the magnitude of these corrections as a function
of the relative difference normalized by $m_{\text{max}}$ between
powers in order to allow a comparison between different system sizes.
While the homogeneous contributions $C_0$ are found not to be the most
dominant, the corrections are peaked at $C_1$ for both systems
considered and decay quickly with $k$. 
In particular, this decay happens increasingly rapidly with larger 
$N$ as a function of relative difference between powers, $k/m_{\text{max}}$ (cf.\ the red curve
marked by squares ($\Box$) for $N=6$ and the black one marked by
diamonds ($\Diamond$) for $N=4$ in
Fig.~\ref{fig:coefficients}) (bottom panel).   
A deviation of the values of $\epsilon_A$ and $\epsilon_B$ from their
value of $\epsilon_c$ is found to reduce the dominance of the peaks presented
above but does not immediately invalidate the assumption.
However, when moving far away from the homogeneous critical point, the
peaks flatten out indicating a breakdown of our approximation.
In summary, all of the above findings can be considered to support the 
assumption about a dominant contribution of homogeneous
powers of recovery rates in Eq.~(\ref{eq:A_0}). 
We have undertaken a similar analysis for the coefficient $A_1$ and expect 
the same behavior for  the remaining $A_n$. 
An analysis of $A_n$ (for $n \ge 2$) 
in a similar manner quickly  becomes prohibitive due to the computational 
complexity of the  resulting expressions. 
However, as $\lambda_1$ approaches zero with increasing $N$, these higher
terms are expected to become increasingly irrelevant.

The question of whether one always expects to obtain a lower bound is
addressed in the discussion (Sec.~\ref{sec:discussion})
after more example cases have been compared to simulation data.

\subsection{\label{sec:disorder_exponents}Critical Exponents from
  Quasi-Stationary Simulations}

Investigations of the $2d$ DCP have been carried out in the past and
have investigated both dynamic \cite{dickman_96} and static scaling
properties of the process \cite{dickman_98}.
In general, the study of critical properties of the disordered process
is complicated due to long relaxation times and ambiguity regarding the
nature of scaling.
In the following, we will investigate the static scaling of the
disordered process by employing QS simulations
\cite{oliveira_2005} and compare our results to both previous studies
as well as theoretical predictions.

In the clean CP, the order parameter, $\lim_{t \to \infty} \rho$ is
expected to obey the scaling form \cite{Marro_99:book}
\begin{eqnarray} 
\rho \sim L^{-x} ~ G\left(L^{1/\nu_{\perp}}(\epsilon-\epsilon_c) 
\right) 
\label{eq:rho_fss} 
\end{eqnarray} 
where $x=\beta/\nu_{\perp}$, $L$ is the linear size of the system, and
$\beta$ and $\nu_{\perp}$ are critical exponents.
Further, $G$ is a scaling function which
asymptotically behaves as $G(y) \to y^{\beta}$ as $y \to \infty$ and
$G(y) \to \text{const.}$ for $y \to 0$.
An analogous finite-size scaling form is expected to be obeyed by the
order parameter fluctuations, $\chi = L^{d} \left( \overline{\rho^2} -
  \overline{\rho}^2 \right)$, with the exponent $x$ replaced by
$x'=-\gamma/\nu_{\perp}$.

In order to apply the above scaling relations, one
commonly considers QS values of observables as no true
stationary state can exist in a finite system.
The CP, when started from a fully infected system, initially
relaxes while spatial correlations grow towards the system size and
temporal correlations decay.
Once the spatial correlation length becomes comparable to the size of
the system, the process enters a QS regime characterized by a
time-independent non-zero transition rate to the absorbing state.
In this regime, the QS density $\overline{\rho}$, i.e. the density
$\rho$ conditioned on survival, attains a constant value.
In the past, analysis of this metastable state in computer simulations
has proved to be notoriously difficult.
Usually, the time-dependent density of infected sites conditioned on
survival, $\overline{\rho}$, which becomes stationary in the QS
regime, is investigated \cite{Marro_99:book}.
Problematically though, it is neither clear at what time 
this density has converged to its QS value nor when the QS state
starts to decay due to finite-size effects \cite{Lubeck_2003}.
Therefore, a range of alternative approaches have been proposed which
enable an observation of this metastable regime (see
Ref.~\cite{oliveira_2005} and references therein).
Here, we employ the QS simulation method \cite{oliveira_2005} which
allows a direct sampling of the QS state by eliminating the absorbing
state and redistributing its probability mass over the active states.

Following Ref.~\cite{oliveira_2005}, one starts from the master equation
Eq.~(\ref{eq:master_equation}).
For the CP, this equation does not admit a non-trivial stationary
solution for a finite system due to the existence of the absorbing
state $\mathbf{0}$ which can be entered but not be left.
The QS solution mentioned above can be defined as
\begin{equation}
\overline{P}(\mathbf{S}) = \lim_{t \to \infty} \frac{P(\mathbf{S},t)}{P_s(t)}~,
\end{equation}
where $P_s(t)$ denotes the survival probability of the process at time $t$.
Now, consider a modification of the governing equation,
\begin{eqnarray}
\partial_t Q(\mathbf{S},t) & = & \sum_{\mathbf{S}'} [  r_{\mathbf{S}'
    \rightarrow \mathbf{S}} Q(\mathbf{S}',t) - r_{\mathbf{S}
    \rightarrow \mathbf{S}'} Q(\mathbf{S},t) \nonumber \\ 
& & + r_{\mathbf{S'} \to  \mathbf{0}} Q(\mathbf{S'},t) Q(\mathbf{S},t) ] ~.
\label{eq:qs_master_equation}
\end{eqnarray}
where $Q(\mathbf{S},t)$ denotes the probability of a new process
governed by this equation being in state $\mathbf{S}$ at time $t$.
The stationary solution of Eq.~(\ref{eq:qs_master_equation}),
$\overline{Q}(\mathbf{S})$, coincides with the QS probability of the
original process as can be seen by substituting $Q(\mathbf{S},t) =
P_s(t) \overline{P}(\mathbf{S})$ and noticing that in the QS regime
$dP_s/dt = -P_s \sum_{\mathbf{S}} r_{\mathbf{S} \to \mathbf{0}}
\overline{P}(\mathbf{S})$.
In that case, the right-hand side of Eq.~(\ref{eq:qs_master_equation})
is equal to zero if $\overline{Q}(\mathbf{S}) =
\overline{P}(\mathbf{S})$ as required.
The last term in Eq.~(\ref{eq:qs_master_equation}) can be viewed as a
redistribution of probability from the absorbing state to the active
states according to their probability \cite{oliveira_2005}.
Thus, if one could sample from a process governed by
Eq.~(\ref{eq:qs_master_equation}), it would converge to a true stationary
state governed by the QS probability distribution of the original
process.
Such a process is given by the original CP where all transitions to
the absorbing state are instead redirected to an active state randomly chosen
according to its probability.
As in practice this probability is not known a priori, an estimate is
generated by sampling from the history of the process.
Generally, the method has proved to be efficient with fast
and reliable convergence after optimization of history sampling
parameters \cite{oliveira_2005,Oliveira2005}.
The approach is particularly suited to a study of the DCP for which,
in dynamic single-seed MC simulations employed for the DCP in the past
\cite{dickman_96,Vojta_05}, the question of whether the asymptotic
limit of the process had been reached was frequently contested.
In contrast, QS simulations offer a clear means of
ensuring this: a true stationary average whose convergence can be
monitored.

\begin{table}[!ht]
\caption{\label{tab:disorder_exponents} 
The critical rates $\epsilon_B$ for a given $\epsilon_A$, critical
exponents $x$ and $x'$ for the
disordered systems discussed in the text.
}
\begin{ruledtabular}
  \begin{tabular}{cccc}
 $\epsilon_A$  & $\epsilon_B$ (critical)& $x$ & $x'$  \\
    \hline
    0.60653   & 0.60653(3) & 0.795(4)  & 0.42(3)  \\
    0.595     & 0.6188(3)  & 0.796(5)  & 0.41(5) \\
    0.5       & 0.7676(4)  & 0.83(1)  & 0.39(3)  \\
    0.4       & 1.1815(5)  & 0.92(4)  & -  \\
    0.35       & 1.7775(5)    & 0.93(4)  & -  \\
    0.3       & 3.89(1)    & 0.99(5)  & -  \\
\end{tabular}
\end{ruledtabular}
\end{table}

Here, we have investigated the $2d$ DCP with bimodal disorder in its
recovery rates drawn from the distribution Eq.~(\ref{eq:bimodal}) by
means of QS simulations for up to $10^8$ time steps and systems 
of sizes $L=8,\dots,128$ sites averaging over no less than $10^3$
disorder realizations.
At the critical point, fits to the above finite-size scaling relations
yielded estimates for the exponents $x$ and
$x'$.
For the homogeneous case, the well-established values for the
exponents of the DP universality class are recovered ($\beta /
\nu_{\perp}=0.795(7)$, $\gamma / \nu_{\perp}=0.41(2)$ \cite{luebeck_04a}).
As the degree of disorder, i.e. 
the difference between recovery rates $\epsilon_A$ and $\epsilon_B$,
 is increased, the measured exponents are found to change with
 disorder strength where $x$ increases while $x'$ decreases
(cf.\ Tab.~\ref{tab:disorder_exponents}).
For strong heterogeneity, no credible fluctuation exponent could be
extracted from the data due to strong sample-to-sample fluctuations.
This is unfortunate as it prevents us from testing the validity of the
hyperscaling relation $\frac{\gamma}{\nu_{\perp}} =
d-\frac{2\beta}{\nu_{\perp}}$.
A similar relation for dynamical exponents had previously
\cite{dickman_96} been found to break down for the DCP.


%
%
\begin{figure*}[ht!]
\flushleft
\subfigure[ ~Lattice (i) ]{
\scalebox{0.2}{\includegraphics[scale=1,angle=270]{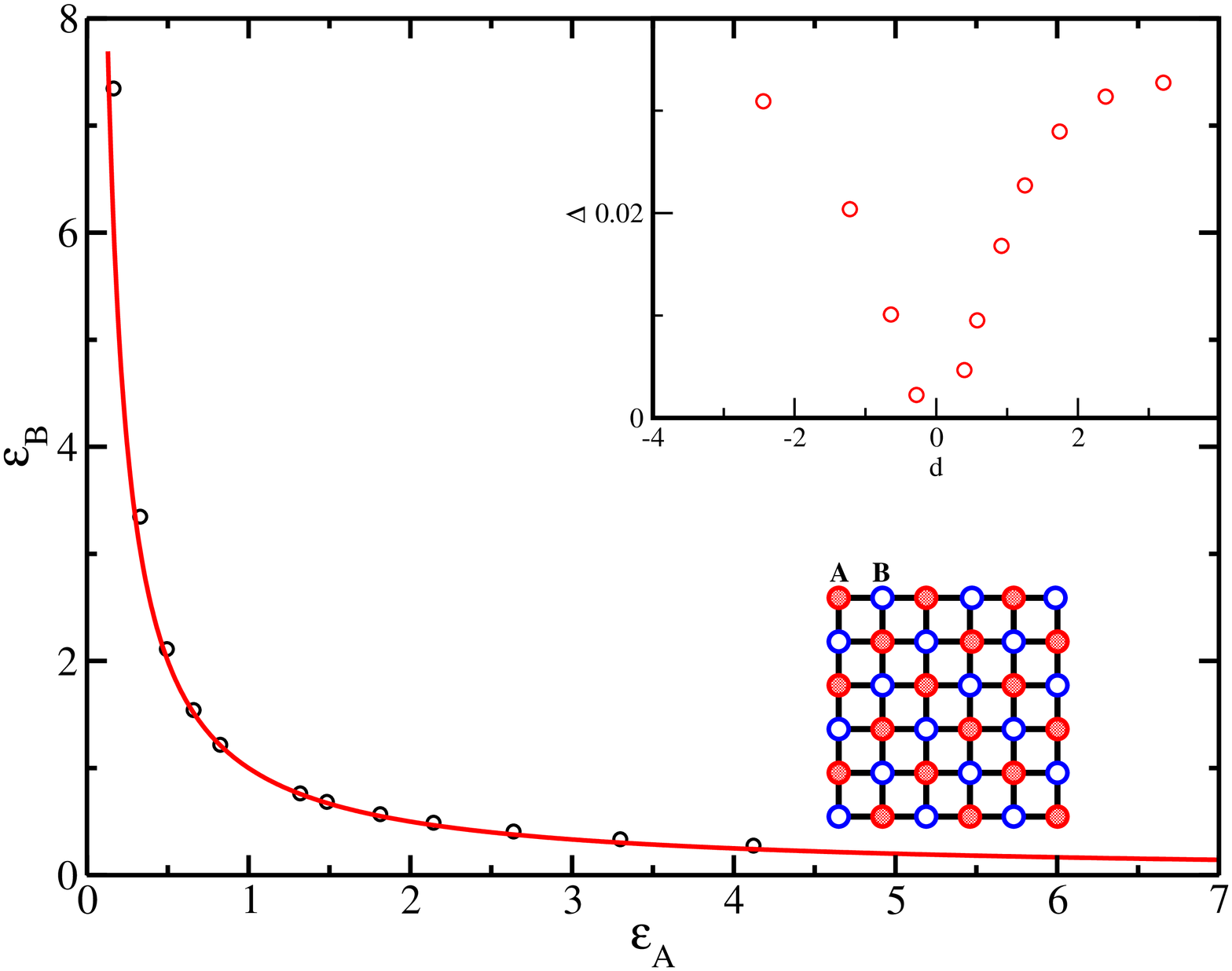}}
\label{f1}
}
\subfigure[ ~Lattice (ii) ]{
\scalebox{0.2}{\includegraphics[scale=1,angle=270]{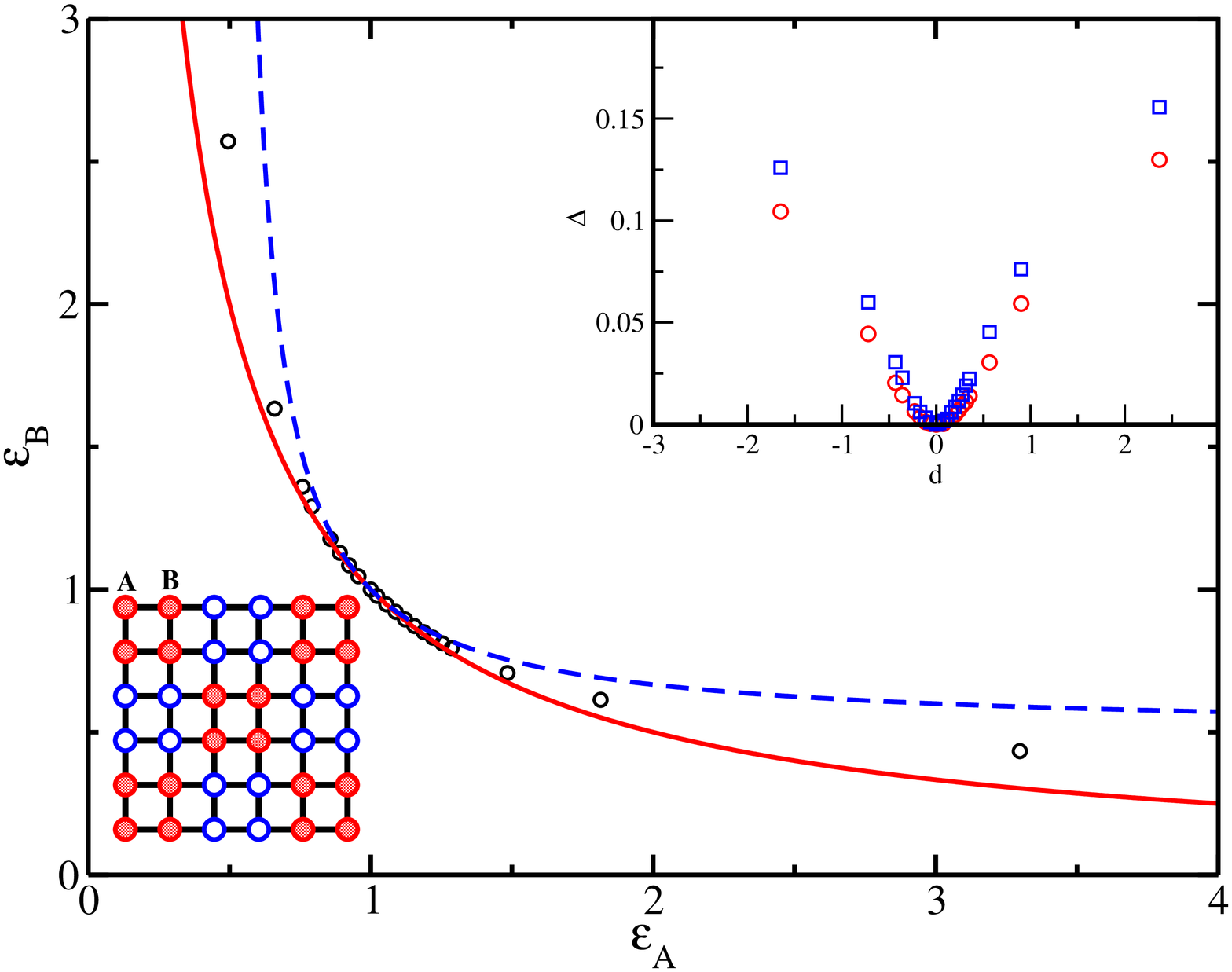}}
\label{f2}
}
\subfigure[ ~Lattice (iii) ]{
\scalebox{0.2}{\includegraphics[scale=1,angle=270]{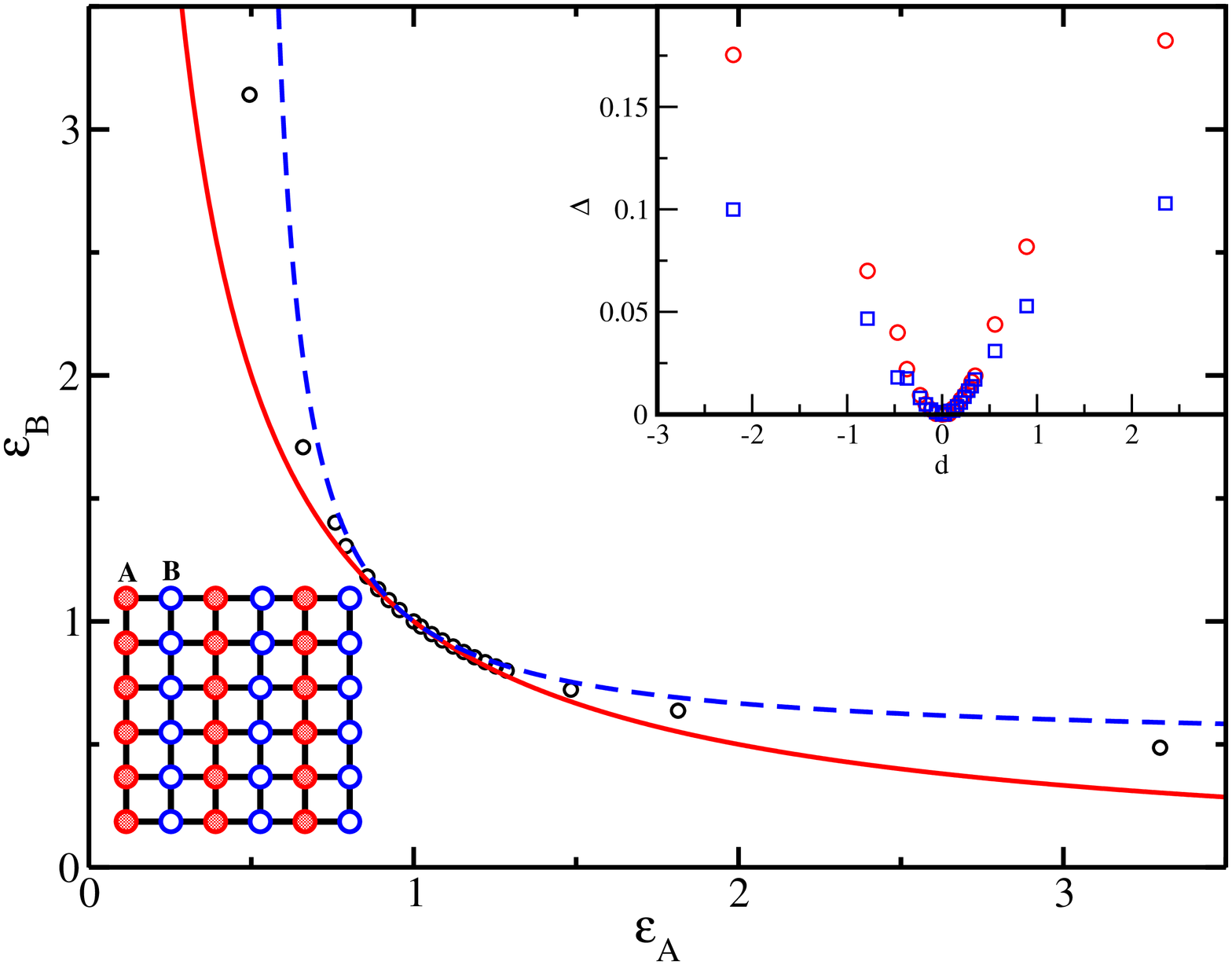}}
\label{f3}
}
\subfigure[ ~Lattice (iv) ]{
\scalebox{0.2}{\includegraphics[scale=1,angle=270]{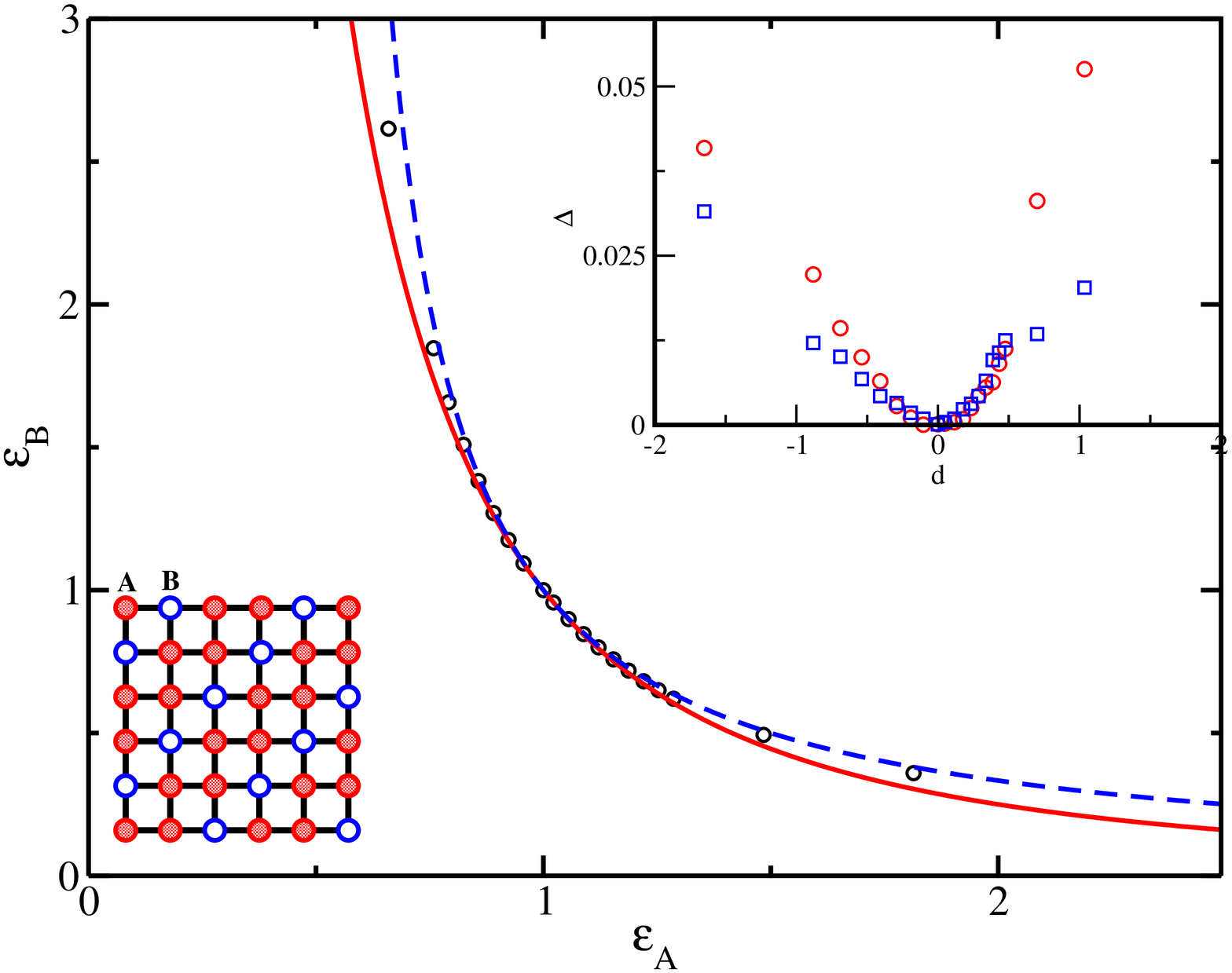}}
\label{f4}
}
\subfigure[ ~Lattice (v) ]{
\scalebox{0.2}{\includegraphics[scale=1,angle=270]{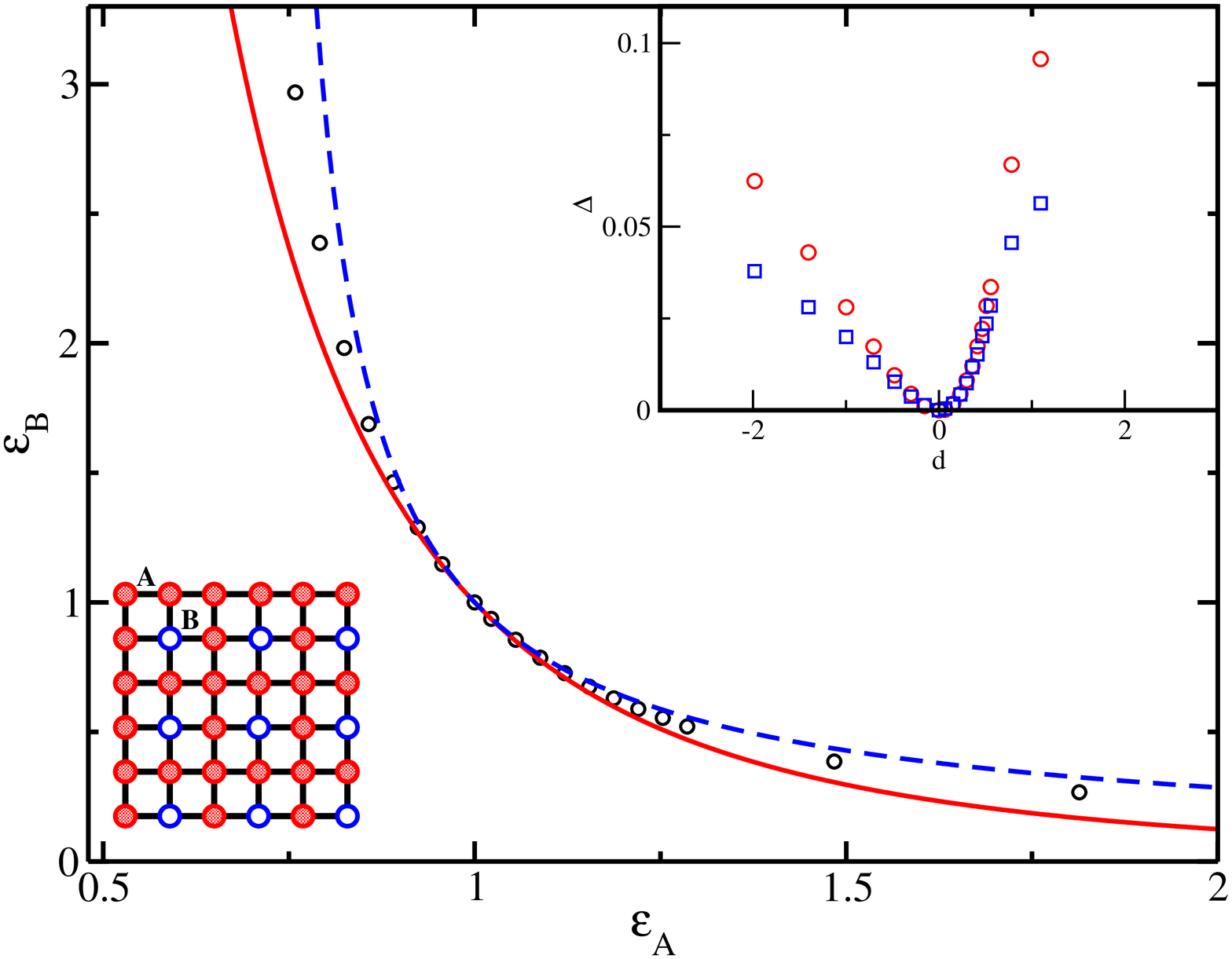}}
\label{f5}
}
\subfigure[]{
\scalebox{0.2}{\includegraphics[scale=1,angle=270]{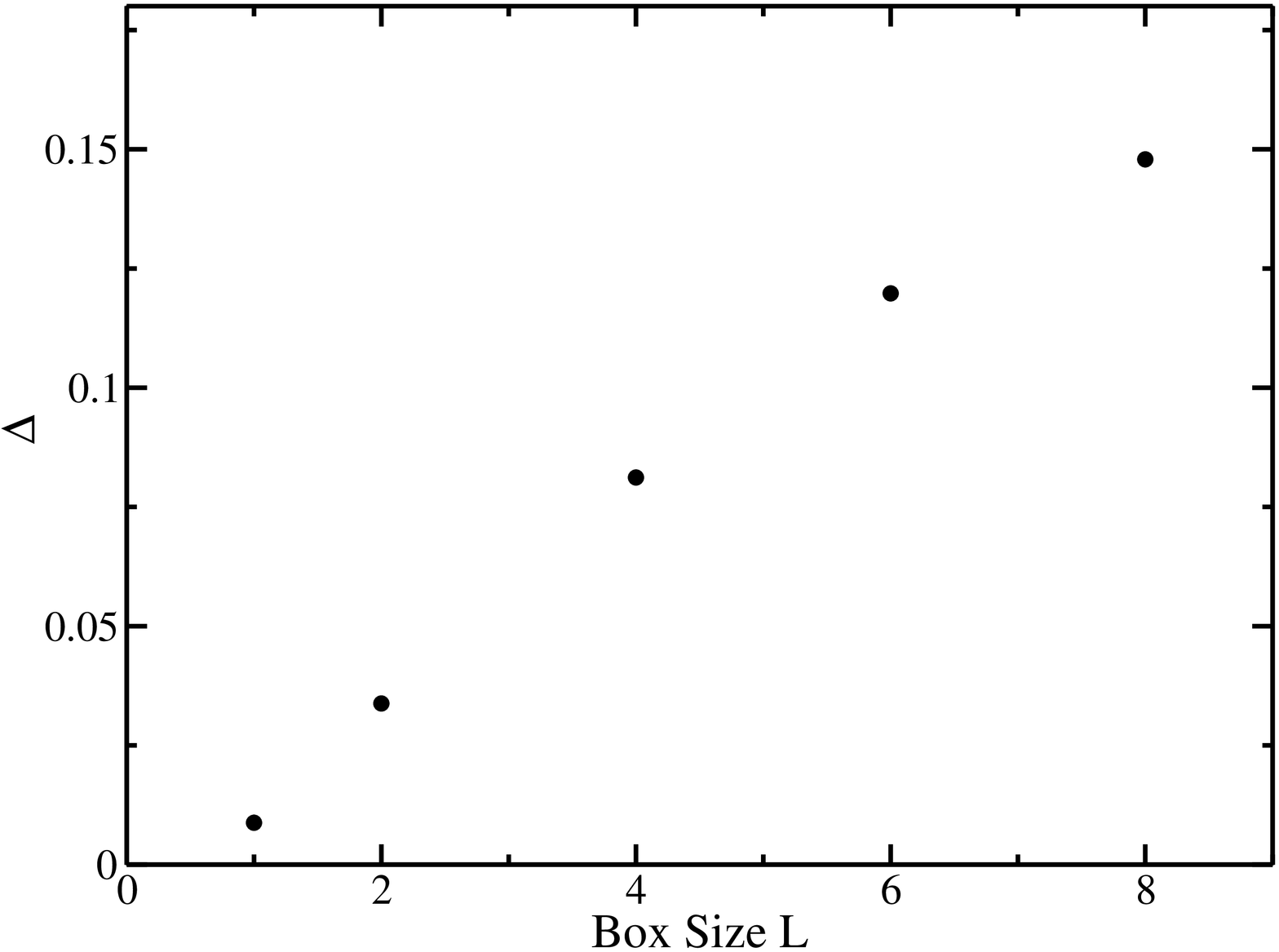}}
\label{fig:box_scaling}
}

\caption[]{(Color online)\\(a)-(e):
 The phase diagram for the CP on the lattices (i)-(v) as
 defined in the text. 
 The circles represent the MC data, the solid
 line is given by Eq.~(\ref{eq:locus_AB}) for the specific lattice, and
 the dashed line corresponds to the mean
 field result taken from Tab.~\ref{t1}. Inset shows the
 deviation $\Delta(d)$ for both approximations (mean field$=\Box$,
 Eq.~(\ref{eq:locus_AB})$=\circ$).\\
 (f): The deviation $\Delta$ as defined in the text at criticality for
 $\epsilon_A=0.9$ for the CP on a lattice such as case (ii) but with
 variable linear size $L$ of the square contiguous regions of $A$ or
 $B$ sites.}
\label{fig:het_phase_diags}
\end{figure*}

\section{\label{sec:hetero} Heterogeneous Periodic Lattices}

In order to investigate the range of validity of Eq.~(\ref{eq:locus}),
we now turn to the behavior of the CP on lattices with periodic
arrangements of sites of type $A$ and $B$ discussed above.

For such systems, the equation for the locus of critical points reads
\begin{equation}
\epsilon_A^{c_A}\epsilon_B^{c_B}=\epsilon_c~, 
\label{eq:locus_AB}
\end{equation}
where $c_A$ and $c_B$ denote the concentration of species $A$ and 
$B$, respectively. 
Three lattice systems have been analyzed with $c_A=c_B=1/2$: 
(i) a standard chessboard lattice [Fig.~\ref{f1}], 
(ii) a big chessboard lattice [Fig.~\ref{f2}] and 
(iii) a lattice of rows [Fig.~\ref{f3}]. 
Extensive MC simulations ($3 \times 10^6$ runs up to $t=3000$ maximum time steps) 
starting from a single infection seed were performed for these
heterogeneous lattices. 
Unlike in the disordered case, for heterogeneous systems, asymptotic
scaling relations that are well-known from the homogeneous case are
found to hold.
At criticality, the average number of infected sites $\langle N(t) \rangle$,  the mean squared radius $\langle R^2 \rangle$ of spread of the CP 
 (where angular brackets denote averaging over all realizations and
 over active realizations at time $t$, respectively) and the survival probability $P(t)$ follow asymptotic scaling laws
 \cite{Marro_99:book},
\begin{equation}
\langle N \rangle \sim t^\eta~,  \quad
\langle R^2 \rangle \sim t^{2/z}~, \quad  P \sim t^\delta~,
\label{eq:dyn_scaling}
\end{equation}
where $\eta$, $\delta$ and $z$ are the dynamical critical exponents
characteristic of the universality class.
These scaling relationships provide a method for 
finding the critical value of the control parameter by fitting the
observables to the above scaling forms following Ref.~\cite{Grassberger_89}.
Furthermore, the dynamical critical exponents can be determined from
the fit.

As expected, the numerical data  agree very
well with the analytical predictions given by Eq.~(\ref{eq:locus_AB}) 
[cf.\ the circles with the solid line for 
$\varepsilon_A \simeq\varepsilon_B \simeq 1 $ in Figs.~\ref{f1}-\ref{f3}]
in the neighborhood of the homogeneous critical point, and start to deviate 
from the predicted phase-separation line for $\varepsilon_{A,B}\agt 1$
consistent with the validity of our approximation. 
The quality of the analytical approximation is high for the 
standard chessboard case ($\Delta < 0.03$ for a very large range of
rates) but becomes worse for the big chessboard ($\Delta$ up to $0.15$
when moving away far from the homogeneous point) and 
especially for rows in the range of large values of $\varepsilon_{A,B}\gg 1$. 

Furthermore, we have studied two lattices with different concentrations of 
nodes A and B, i.e. $c_A/c_B = 2/1$ -- lattice (iv) [see Fig.~\ref{f4}], 
and $c_A/c_B = 3/1$ -- lattice (v) [see Fig.~\ref{f5}]. 
In these cases, the phase-separation lines are not symmetric about 
the bisector in the $\varepsilon_A-\varepsilon_B$ plane. 
As can be seen from Figs.~\ref{f4}-\ref{f5}, the results of MC 
simulations of the CP on these lattices are again in good
agreement with the analytical expression given by
Eq.~(\ref{eq:locus_AB}), 
especially near the homogeneous critical point (cf.\ the circles with
the solid line for $\varepsilon_A\simeq\varepsilon_B \simeq 1 $ in
Figs.~\ref{f4}-\ref{f5}).
In case of  lattice (iv), the error as shown in the inset indicates a 
similar order of magnitude degree of accuracy as in the simple
chessboard case before ($\Delta < 0.05$ for a
large range of rates) while for lattice (v) the approximation is found
to deteriorate [with approximately twice the value of $\Delta$ as
compared to lattice (iv)].

It is instructive to compare the expression for the phase-separation line
given by Eq.~(\ref{eq:locus_AB}) with the results obtained from the
master equation 
within the standard mean field approximation. 
Expressions similar to Eqs.~(\ref{eq:disorder_mf}) can be found for all
the different lattices and solved for the critical rates in the
steady-state regime.
The resulting expressions for all the lattices are summarized in
Table~\ref{t1}.

\begin{table}
\caption{\label{t1} 
The expressions for the phase-separation lines for the CP on different
lattices (first column) obtained according to Eq.~(\ref{eq:locus_AB})
(second column) and within the standard mean field approach (third column). 
}
\begin{ruledtabular}
  \begin{tabular}{ccccc}
      ~     & lattice type & Eq.~(\ref{eq:locus_AB})   & mean field  \\
    \hline
  &  (i)   & $\varepsilon_B=1/\varepsilon_A$  & $\varepsilon_B=1/\varepsilon_A$  \\
   &  (ii)  & $\varepsilon_B=1/\varepsilon_A$  & $\varepsilon_B =
    \varepsilon_A/(2\varepsilon_A - 1)$ \\ 
   &  (iii)  & $\varepsilon_B=1/\varepsilon_A$  & $\varepsilon_B =
    \varepsilon_A/(2\varepsilon_A - 1)$ \\ 
   &  (iv)  & $\varepsilon_B=1/\varepsilon_A^2$  & $\varepsilon_B =
    1/(2\varepsilon_A - 1)$ \\ 
    & (v)  & $\varepsilon_B=1/\varepsilon_A^3$  & $\varepsilon_B =
    \varepsilon_A/(2\varepsilon_A^2 - 1)$ \\ 
\end{tabular}
\end{ruledtabular}
\end{table}

As follows from Table~\ref{t1}, the mean field result coincides with the
expression for the phase-separation line given by
Eq.~(\ref{eq:locus_AB}) 
for the
standard chessboard configuration [lattice (i)] and gives a different
prediction for all other cases studied.
The rescaled mean field  results agree very well with 
MC data around the homogeneous point but 
display deviations for $\varepsilon_{A,B}\gg 1$. 
Looking at the corresponding errors $\Delta(d)$, they are found to be
of the same order as found for the previous analytical approximation.

The fact that for the simple chessboard lattice our earlier prediction
and the mean field result coincide reveals this case to be special in that the
rescaled mean field does not over- but underestimate the true critical values. 
In  all other studied lattices, the rescaled mean field results for the 
phase-separation lines lie above
the numerical data [cf.\ the dashed lines with the circles in
Figs.~\ref{f2}-\ref{f5}] and thus lead to an  overestimate of the
value if $\varepsilon_B$ for a given $\varepsilon_A$.
This means that for these cases, mean field estimates of critical values
can serve as an upper bound on the critical recovery rate.
In contrast, the phase-separation lines predicted by Eq.~(\ref{eq:locus_AB}) 
provide a
consistent underestimate of the true critical line for all studied lattices and
therefore a lower bound (cf.\ the solid lines in relation to the circles in 
Figs.~\ref{f2}-\ref{f5} and see the arguments given in
Sec.~\ref{sec:discussion}) for the critical thresholds. 

In order to more systematically investigate how our alternative
analytical approximation deteriorates as the spatial arrangement of
sites becomes ``less mixed'', we consider a lattice like the big
chessboard, lattice (ii), but vary the linear size $L$ of the square contiguous
regions of $A$ or $B$ sites.
The resulting deviation $\Delta(d)$ as defined above at critical
$\epsilon_B$ for a choice of $\epsilon_A=0.9$, i.e. appreciably far 
away from the homogeneous
critical point, as a function of $L$ is shown in Fig.~\ref{fig:box_scaling}.
One can see that for $L \ge 4$ the accuracy quickly becomes worse than 
for any rate and lattice previously considered indicating a rapid breakdown of
the approximation.

Finally, the universality of the critical behavior of 
the CP in binary lattices was investigated. 
The expected dynamical 
power-law scaling relations [see Eqs.~(\ref{eq:dyn_scaling})] were
verified and used to obtain the resulting critical exponents for 
several sets of parameters $(\varepsilon_A,\varepsilon_B)$ on
the phase-separation lines for all the lattices. 
The evaluation of the exponents was performed following 
Ref.~\cite{Grassberger_79} through extensive numerical 
simulations performing averages of $3\times 10^6 - 10^7$ runs to a maximum of 
$t=3000$ time steps.
Our results obtained for the different lattices 
indicate that, within error bars, the exponents in all cases coincide 
with those established for $2d$ processes in the DP universality class
($\eta = 0.2295(10)$, $\delta=0.4505(10)$,$2/z=1.1325(10)$) \cite{Voigt_97}.  
Furthermore, the static scaling exponent ratios determined analogously to
the disordered case from the QS simulation method are
found to coincide with those of the DP universality class.


\section{\label{sec:discussion} Discussion}

Looking back at the phase diagrams for both the disordered and the
periodic systems, the introduction of disorder in the form of a random
placement of $A$ and $B$ sites appears to enhance the activity of the system.
In Fig.~\ref{fig:disorder}, MC data for the disordered system are 
presented and one notices a shallow initial increase in critical
$\varepsilon_B$ for a given $\varepsilon_A$ as one moves away from the
homogeneous  critical point followed by an increasingly steep increase
at values of $\varepsilon_A \alt 0.5$ ($\epsilon_A \alt 0.3$).
Comparing this behavior with the corresponding periodic system
[Fig.~\ref{f2}], the critical value for $\varepsilon_B$ in the 
disordered system is found to be much larger.

Considering the arrangement of, say, $A$ sites as a site percolation
problem, one notices that for concentrations below the percolation
threshold ($x_c \simeq 0.59$) no infinite cluster of such sites can
exist.
Therefore, no matter how small (but non-zero) the corresponding recovery rate
$\varepsilon_A$ is, it will require a finite value of $\varepsilon_B$
to render the system critical as a finite cluster cannot support an
active state indefinitely.
Conversely, above the percolation threshold there exists a finite
value of $\varepsilon_A$ below which the system will be active
irrespective of the value of $\varepsilon_B$.
Therefore, for the case of $x=0.5$, no asymptote at any non-zero value
of $\epsilon_A$ would be expected.
Interestingly, the mean field expression Eq.~(\ref{eq:mf_disorder}) does
predict an asymptote at $\epsilon_A=x$ albeit for any concentration
of sites.

Turning to the CP in heterogeneous periodic lattices, for a range of
cases the combination of the standard mean field approximation and our
alternative analytical approximation is useful in
practice to pinpoint the location of the transition a priori.
Indeed, a tight fit for all cases (with the exception of the
simple chessboard lattice) can be attested.
In particular, the influence of spatial structure on the quality of
our approximation becomes evident.
The less mixed the arrangement of $A$ and $B$ sites becomes, the worse
the fit of the approximation is found to be as indicated by the
results in Fig.~\ref{fig:box_scaling}.
In order to evaluate the practical relevance of our approximation, 
the question of whether it is expected to always yield a lower bound
has to be addressed.
To this end, we define an average clustering coefficient specific
to a particular lattice configuration and site type. 
For sites of type $A$ for instance, define $C_{A}=n_{\text{NN,A}}/Z$ where
$n_{\text{NN,A}}$ denotes the number of nearest
neighbors of site type $A$ (and analogously for $B$ sites).
For the case of a periodic lattice with a $1:1$ mixture of $A$ and $B$
sites, consider the minimally-clustered configuration, that is the
standard chessboard [lattice (i)], for which $C_{A}=C_{B}=0$.
We know that for this case our approximation yields a very tight
lower bound to the true curve of critical points.
Any lattice with the same concentration of sites will necessarily have
a higher clustering coefficient, i.e. a larger fraction of contiguous
regions of $A$ and $B$ sites.
Assuming different recovery rates for the two types of site, the
disease will have the tendency to survive longer in a constellation
$AABB$ as compared to $ABAB$ due to the adjacency of two sites of
lower recovery rate (say $A$) which enhances the probability of
infection and reinfection in the $AA$ arrangement.
This activity-enhancing effect is not offset by the fact that two less
reactive sites (say $B$) are also bordering as their faster (than A
sites) recovery is largely independent of spatial arrangement. 
Indeed, a direct diagonalization of the corresponding Liouville
operator for these two different arrangements of
sites readily confirms this intuition.
One obtains a lower (absolute)
value for the real part of the first non-trivial eigenvalue in case of
an arrangement $AABB$ as compared to $ABAB$ indicating a slower
approach to the absorbing state in a finite system.

From this we conclude that for any periodic arrangement of $A$ and $B$
sites our alternative analytical approximation is expected to yield a
lower bound to the phase separation line.
Similarly, the arrangement used in lattice (iv) is the
minimally-clustered 
($C_A=1/2$, $C_B=0$) arrangement with a $1:2$
concentration of sites and is found to give a lower bound leading us to expect
the same behavior for any arrangement of $A$ and $B$ sites in this
ratio.
Therefore, by testing the minimally-clustered case for the desired
concentration one should in practice be able to verify whether or not a lower
bound is expected by our approximation.

Considering the critical exponents obtained for both the disordered
and the periodic systems, they are found to be disorder dependent in
the former case while they remain at their DP values in the latter.
Predictions from a numerical implementation of the strong-disorder
renormalization scheme in $2d$ predict an exponent value
$x_{\text{strong}}=1.0$ at an IRFP \cite{hooyberghs_04,Motrunich_00}.
At the same time, as conjectured in \cite{hooyberghs_04} and supported
by numerical evidence, the DCP in $2d$ is likely to be dominated by
such a fixed point for sufficiently strong disorder similar to the
$1d$ case.
Thus, one would expect disorder-dependent varying exponents which
approach their values expected at an IRFP
for strong disorder.
Our results support this picture: we find continuously varying
disorder-dependent exponents and a finite-size scaling exponent $x$
which is compatible with the predicted value at an IRFP for
the strongest disorder under consideration ($\epsilon_A=0.3$).

Regarding the unchanged exponents in the case of periodic systems, 
these findings confirm theoretical arguments
\cite{vojta_06_review} which make a prediction about the universal
behavior of the CP in heterogeneous and disordered systems.
Under coarse-graining the heterogeneity present in systems such as
the heterogeneous periodic lattices considered in this paper will 
eventually become homogeneous after a
finite number of iterations of the coarse-graining procedure.
Thus, one would expect the critical behavior of the CP to be governed
by the conventional clean fixed point of the renormalization group
transformations \cite{Odor_04:review}.


\section{\label{sec:conclusion} Conclusion}

In conclusion, we have 
investigated the contact process in both heterogeneous disordered and
periodic $2d$ systems (binary lattices).
The phase diagram has been obtained via extensive Monte Carlo
simulation.
Furthermore, two approximations have been successfully used in order
to constrain a region of phase space which contains the line of
critical points.
First, the mean field approximation was employed to give a phase separation
line which provided an upper bound to this region in almost all systems.
Second, an alternative analytical approximation based on the structure
of the Liouville operator was motivated and used to obtain a
respective lower bound in all cases.
The quality of both approximations was quantitatively analyzed for all
systems and found to be high in the vicinity of
the homogeneous critical point but increasingly worse when moving to
higher degrees of heterogeneity.
In general, we conclude that the strategy of constraining a region
deemed to contain the critical points a priori may be of practical 
interest particularly in
connection with disordered systems in which long relaxation times
render computer simulations very costly.

Lastly, critical exponents obtained for the disordered system are in
good agreement with data from previous investigations obtained in the
crossover region between the homogeneous case and strong disorder.
In particular, the values obtained for the critical exponent $x$ are
compatible with the existence of an IRFP in the $2d$ DCP for
sufficiently strong disorder.
At the same time, as expected the well-known DP exponents were
recovered for all periodic systems.

\section*{Acknowledgments} The computations were
mostly performed on the Cambridge University Condor Grid. SVF and CJN
would like to thank the EPSRC and the Cambridge European Trust for 
financial support.

\bibliographystyle{apsrev}


\end{document}